\documentclass[%
 reprint,
superscriptaddress,
 amsmath,amssymb,
 aps,
]{revtex4-2}

\usepackage{graphicx}
\usepackage{dcolumn}
\usepackage{amssymb}
\usepackage{amsmath,bm}
\usepackage{graphicx,color}
\usepackage{bbm}
\usepackage{newtxmath}
\usepackage{multirow}
\usepackage{xcolor}
\usepackage{hyperref}
\usepackage{ulem}
\usepackage{bm}

\begin{document}

\title{The role of non-affine deformations in the elastic behavior of the cellular vertex model}%

\author{Michael F. Staddon}
\affiliation{Center for Systems Biology Dresden, Dresden, Germany}
\affiliation{Max Planck Institute for the Physics of Complex Systems, Dresden, Germany}
\affiliation{Max Planck Institute of Molecular Cell Biology and Genetics, Dresden, Germany}

\author{Arthur Hernandez}
\affiliation{Department of Physics, University of California Santa Barbara, Santa Barbara, CA 93106}

\author{Mark J. Bowick}
\email{bowick@kitp.ucsb.edu}
\affiliation{Kavli Institute for Theoretical Physics, University of California Santa Barbara, Santa Barbara, CA 93106}

\author{M. Cristina Marchetti}
\email{cmarchetti@ucsb.edu}
\affiliation{Department of Physics, University of California Santa Barbara, Santa Barbara, CA 93106}

\author{Michael Moshe}
\email{michael.moshe@mail.huji.ac.il}
\affiliation{Racah Institute of Physics, The Hebrew University of Jerusalem, Jerusalem, Israel 91904}

\date{\today}

\begin{abstract}
The vertex model of epithelia describes the apical surface of a tissue as a tiling of polygonal cells, with a mechanical energy governed by deviations in cell shape from  preferred, or target, area, $A_0$, and perimeter, $P_0$. The model exhibits a rigidity transition driven by geometric incompatibility as tuned by the target shape index, $p_0 = P_0 / \sqrt{A_0}$. For $p_0 > p_*(6) = \sqrt{8 \sqrt{3}} \approx 3.72$, with $p_*(6)$ the perimeter of a regular hexagon of unit area, a cell can simultaneously attain both the preferred area and preferred perimeter. As a result, the tissue is in a mechanically soft compatible state, with zero shear and Young's moduli. For $p_0 < p_*(6)$, it is geometrically impossible for any cell to realize the preferred area and perimeter simultaneously, and the tissue is in an incompatible rigid solid state. Using a mean-field approach, we present a complete analytical calculation of the linear elastic moduli of an ordered vertex model. We analyze a relaxation step that includes non-affine deformations, leading to a softer response than previously reported. The origin of the vanishing shear and Young’s moduli in the compatible state is the presence of zero-energy deformations of cell shape. The bulk modulus exhibits a jump discontinuity at the transition and can be lower in the rigid state than in the fluid-like state. The Poisson's ratio can become negative which lowers the bulk and Young's moduli. Our work provides a unified treatment of linear elasticity for the vertex model and demonstrates that this linear response is protocol-dependent.
\end{abstract}

\maketitle
\maketitle
\maketitle

\begin{figure}
    \centering
    
  \includegraphics[width=0.5\textwidth]{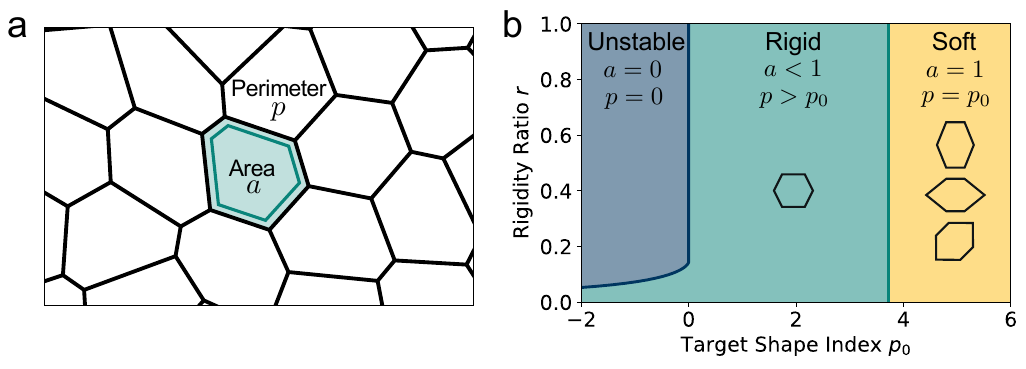}
    \caption{The vertex model for epithelia. (a)  
    (a) The apical surface of an epithelium is modeled by a polygonal tiling, with each polygon representing a cell. (b) Vertex model phase diagram in the $p_0$, $r$ plane. Within the blue region cells are unstable and collapse. Within the green region the tissue is in an incompatible state, with neither preferred perimeter nor area achieved, and the ground state is a regular hexagonal lattice. The tissue acts like a solid in response to shear. Within the yellow region both preferred perimeter and area are achieved and cells have a degenerate ground state and the tissue has zero shear modulus. The cell shapes show example energy minima, where a cell may elongate, increase its pointiness using the angle $\phi$, or increasing its shear tilt angle $\theta$ as described in Ref.~\cite{hernandez2022anomalous}, in order to increase its perimeter while maintaining unit area.
    }
    \label{fig:1}
\end{figure}

Many biological processes, such as morphogenesis~\cite{martin2009pulsed, etournay2015interplay, streichan2018global, maniou2021hindbrain}, wound healing~\cite{poujade2007collective, brugues2014forces, tetley2019tissue, ajeti2019wound}, and cancer metastasis~\cite{friedl2009collective, arwert2012epithelial}, require  coordinated motion and shape changes of many cells. An important open question in biology is how the large scale mechanics of biological tissue emerges from the properties of individual cells, which are in turn governed by force-generating proteins within the cytoskeleton and adhesion molecules between cells~\cite{salbreux2012actin, murrell2015forcing, ladoux2017mechanobiology}. Many theoretical models have been proposed to describe dense epithelia, single layers of very tightly packed cells~\cite{graner1992simulation, marchetti2013hydrodynamics, banerjee2019continuum, alert2020physical, alt2017vertex}. Among these, vertex models, originally developed from models of soap films~\cite{nagai2001dynamic}, have proven a powerful starting point for capturing the mechanical properties of epithelia. The vertex model describes the apical surface of a confluent tissue as a polygonal tiling of the plane (Fig.~\ref{fig:1}a)~\cite{alt2017vertex, nagai2001dynamic, fletcher2014vertex, farhadifar2007influence, staple2010mechanics, bi2016motility}. Each polygon represents a cell, each edge a cell-cell junction, and each vertex a multicellular junction. Each cell’s mechanics are controlled by multiple bio-mechanical processes that were proposed to be effectively described by a mechanical energy determined by deviations of their area and perimeter from preferred values. These preferred values encode bio-mechanical properties such as cadherin molecules concentration, apical ring-contractiliy and more.  Force balance via energy minimization then determines the position of the vertices and thus the shape of cells in the tissue. Topological rearrangements resulting in cell intercalation, cell division and motility have also been incorporated, and the model  has been highly successful in capturing a range of biological processes, such as tissue growth~\cite{hufnagel2007mechanism}, wound healing~\cite{tetley2019tissue}, and tissue organization~\cite{farhadifar2007influence}.

It has been shown that  vertex models exhibit a transition between a fluid-like state and a solid-like state where cells are jammed and unable to rearrange (Fig.~\ref{fig:1}b). This rigidity transition occurs at constant cell density and is driven by both active processes, such as  fluctuations in cell-edge tension and cell motility~\cite{bi2015density, bi2016motility, sussman2018anomalous,  tong2021linear, duclut2021nonlinear, duclut2022active,farhadifar2007influence, staple2010mechanics}, as well as by geometric constraints~\cite{moshe2018geometric,hernandez2022anomalous}. Recent work by us and others has shown that even in the absence of fluctuations and topological rearrangements, vertex models exhibit a rigidity transition associated with geometrical frustration~\cite{moshe2018geometric,merkel2019minimal,hernandez2022anomalous}.  In the rigid or incompatible state cells are unable to achieve the target values of area and perimeter and the system is under finite tension, with a unique gapped ground state. In the soft or compatible state, cells achieve both target area and perimeter and the ground state has zero energy. Due to the underconstrained nature of the vertex model, however, the liquid ground state is degenerate as for a given $n$-sided polygon there are many shapes that preserve area and perimeter. This allows the system to accommodate small shear deformations by finding  a new zero energy shape, resulting in vanishing shear modulus.

The linear elastic response of an ordered hexagonal vertex model to external deformations has been examined through calculations of the shear and bulk moduli~\cite{staple2010mechanics, murisic2015discrete,hernandez2022anomalous}. Staple \emph{et al.}~\cite{staple2010mechanics,hernandez2022anomalous} evaluated the elastic moduli and first demonstrated the vanishing of the shear modulus in the compatible state. More recently, we showed that the vanishing of both shear and Young moduli in the soft regime stems from the degeneracy of the compatible ground states, which allows the deformed tissue to spontaneously shear to a new compatible ground state to accommodate the external deformation~\cite{hernandez2022anomalous}. We additionally discovered that the response  is highly singular at the critical point, with breakdown of linear elasticity and anomalous coupling between compression and shear, as quantified by the development of a new elastic constant~\cite{hernandez2022anomalous}. 

The above studies  only allow for affine deformations of the cells. This approximation can be viewed as appropriate for determining the short-time response of the vertex model to strain. The vertex model has, however, additional degrees of freedom and can relax stress by moving vertices in a non-affine way. Murisic \emph{et al.}~\cite{murisic2015discrete} incorporated these effects by considering the hexagonal lattice as the union of two sub-lattices with a microscopic shift between them and found  that the shear modulus is $2/3$ softer than previously reported. Tong \emph{et al.}~\cite{tong2021linear} used simulations to measure the shear storage modulus and viscosity in both ordered and disordered model tissues.

In this paper, we expand upon previous work  by incorporating simple \textit{non-affine} deformations. Using a mean field model for a hexagonal lattice, we derive analytic expressions for all the linear elastic moduli of the tissue, and verify these results using simulations. We show that, away from the critical point, the elastic constants of a regular VM satisfy the standard relations of two-dimensional elasticity of isotropic solids. Despite this Hookean relationship, the mechanical linear response exhibits robust non-affine contributions that can significantly reduce the elastic constants, as known to happen in amorphous solids~\cite{falk1998dynamics, utter2008experimental, ellenbroek2009non, zaccone2011approximate}. For instance, the bulk modulus can be softer in the rigid state than in the soft fluid-like state and jumps discontinuously  across  the solid to fluid transition. We highlight several novel behaviors of vertex model elasticity, such as negative Poisson's ratio and a softening of the tissue as the ratio of area to perimeter stiffness increases. We verify our analytical results using numerical vertex model simulations of a regular tissue. 

The remainder of the paper is organized as follows. In Section \ref{sec:modelandsimulation} we state the vertex model simulation and deformation protocol to extract various elastic constants. In Section \ref{sec:MFT} we introduce the VM and its mean-field implementation used in the present work, and present a new derivation of the ground states that allows us to quantify the degeneracy of the compatible regime.  In Section \ref{sec:mechanics}, after  highlighting the distinction between the affine and non-affine deformations allowed in our model, we present results for all the elastic constants. We conclude in Section \ref{sec:conclusions} with a brief discussion.
\section{Vertex model: simulation and deformation protocol}
\label{sec:modelandsimulation}
\subsection{The vertex model of epithelia}
The vertex model describes cells in a confluent tissue as polygons of area $A_\alpha$ and perimeter $P_\alpha$ (Fig.~\ref{fig:1}a). The tissue  energy is written as
\begin{equation}
    E_{\text{tissue}} = \frac{1}{2}\sum_\alpha K(A_\alpha - A_{\alpha 0})^2 + \frac{1}{2} \sum_\alpha \Gamma P_\alpha^2 + \sum_{\langle ij\rangle} \Lambda_{ij} L_{ij},
\end{equation}
where $\alpha$ labels individual cells and $\langle ij\rangle$ indexes edges connecting vertices $i$ and $j$. The first term  embodies the energy cost of  cell area deformations, with $K$  the area elasticity and $A_{\alpha 0}$ the preferred or target area. The second term represents active contractility and elasticity of the cytoskeleton, with $\Gamma$ the contractility. The third term represents interfacial energy between neighboring cells, with $L_{ij}$ the length of edge $ij$ and  $\Lambda_{ij}$ the associated tension controlled by the interplay of cell-cell adhesion and cortex contractility. The tension can become negative when adhesion overcomes contractile surface forces.

The mechanical force on vertex $i$ with position $\mathbf{x}_i$ is given by $\mathbf{F}_i = -\frac{\partial E_\text{tissue}}{\partial \mathbf{x}_i}$. The tissue rearranges vertices to locally minimize the energy. This can be described quasi-statically by requiring force balance at each time-step, or dynamically by assuming that vertices relax according to overdamped dynamics where viscous drag balances the mechanical forces: $\gamma \frac{\partial \mathbf{x}_i}{\partial t} = \mathbf{F}_i$, with $\gamma$ a friction coefficient.

As the network relaxes, edges may shorten and cells may shrink, resulting in topological rearrangements that reconfigure the network. In T1 transitions, also known as cell-cell intercalations, a junction between two cells shrinks to a point and a new edge is formed, causing two originally neighboring cells to lose contact and two previously unconnected cells to form a new interface. T1 transitions allow the tissue to relax shear stresses through cell rearrangements rather than cell elongation. A T2 transition, also known as cell extrusion, occurs as a cell shrinks to zero area and is replaced by a single vertex. The mechanical state of the tissue is controlled by both topological rearrangements driven by active processes and geometric frustration. Both types of processes can drive transitions between rigid and fluid states. Here we neglect topological rearrangements to focus on the role of geometry.


We further simplify the model by assuming that all cells have the same preferred area $A_{\alpha 0} = A_0$ and all edges have the same tension $\Lambda_{ij} = \Lambda$. The interfacial energy can then be written in terms of the cell perimeter,  $\sum_{\langle ij\rangle} \Lambda L_{ij} = \frac{1}{2} \sum_\alpha \Lambda P_\alpha$, where the factor of $\frac{1}{2}$ arises because the interfacial energy of each edge is shared by two cells. The tissue energy can then be recast in the form
\begin{equation}
    E_{\text{tissue}} = \frac{1}{2}\sum_\alpha K(A_\alpha - A_0)^2 + \frac{1}{2} \sum_\alpha \Gamma (P_\alpha - P_0)^2 + E_0\;,
\end{equation}
where $P_0 = - \frac{\Lambda}{2 \Gamma}$ is the preferred perimeter, and $E_0$ is a constant term obtained from completing the square. Since we care about the gradient of energy and not the absolute value, we discard $E_0$ in the following.

Finally, we work in dimensionless units by normalizing the energy with $K A_0^2$ and lengths with $\sqrt{A_0}$. The dimensionless tissue energy is then given by
\begin{equation}
   E_{\text{tissue}} = \frac{1}{2}\sum_\alpha(a_\alpha - 1)^2 + \frac{1}{2}\sum_\alpha r(p_\alpha - p_0)^2,
\end{equation}
where $a_\alpha = A_\alpha / A_0$, $p_\alpha = P_\alpha /\sqrt{A_0}$, $r = \Gamma / K A_0$ is the rigidity ratio, and $p_0 = P_0 / \sqrt{A_0}$ is the target shape index of the cell.

\subsection{Deformation Protocol}

To numerically obtain the elastic moduli, we simulate the mechanical response of the vertex model under different deformations using a tissue of 4 hexagonal cells in a periodic box of lengths $L_x(0)$ and $L_y(0)$, and area $A(0) = L_x(0) L_y(0)$ determined by energy minimisation, and implemented in the Surface Evolver software~\cite{brakke1992surface}. For the incompatible regime, the ground state is a regular hexagonal cell. For the compatible regime, while the ground state is degenerate, we use a hexagon with $120^\circ$ angles between edges and with the edge lengths determined by energy minimisation. First, we use an intermediate rigidity ratio of $r = 0.1$, and test the response across a range of preferred values of the shape index, from $p_0 = 0$ to $p_0 = 4.6$, covering both the compatible and incompatible regimes.

\begin{figure}
    \centering
    \includegraphics[width=0.5\textwidth]{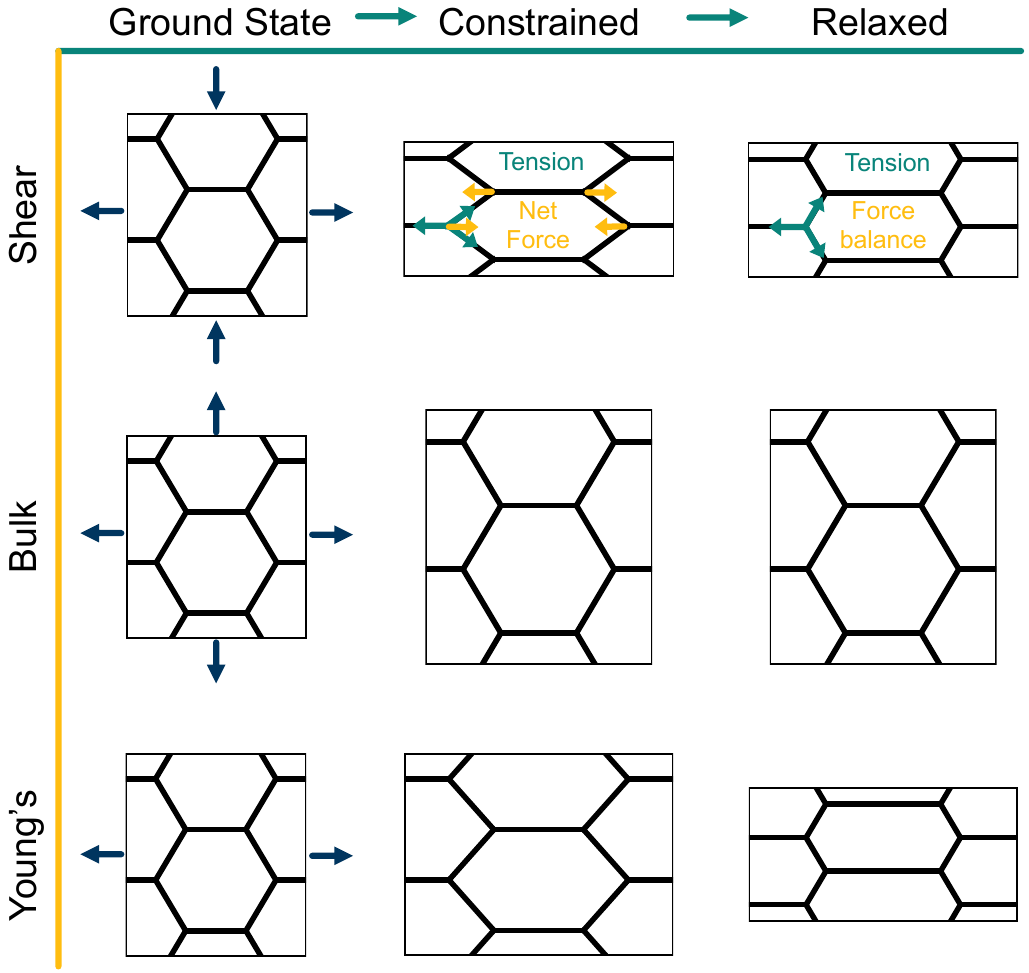}
    \caption{ Strain protocols for measuring elastic moduli of the vertex model. (Top, Middle, Bottom) From the ground state, the periodic box lengths and vertex positions are transformed and constrained according to an affine transformation, shown by the arrows. From the constrained state, the system is relaxed according to tissue-scale or box constrained. (Top) The shear modulus is calculated by applying a shear transformation to the box. In the constrained state, every edge has the same tension, producing a net force on the vertices, hence this is not a force-balanced state. After relaxation, forces are balanced through a non-affine transformation on the vertices. During relaxation the box size is fixed. (Middle) The bulk modulus is calculated by applying an isotropic expansion to the box and vertices. During relaxation the box size is fixed. (Bottom) The Young's modulus and Poisson's ratio are calculated by applying a uniaxial strain to the box and vertices. During relaxation the height of the box may change and vertices may move.}
    \label{fig:2}
\end{figure}

To calculate the shear modulus, we deform the ground state (Fig.~\ref{fig:2} top-left) by applying an initially affine deformation to vertices and the boundaries: $x_i(\epsilon) = (1 + \epsilon / 2) x_i(0) $, $y_i(\epsilon) = (1 + \epsilon / 2)^{-1} y_i(0) $, and $L_x(\epsilon) = (1 + \epsilon / 2) L_x(0)$ and $L_y(\epsilon) = (1 + \epsilon / 2)^{-1} L_y(0)$, where $\epsilon = 0.001$ (Fig.~\ref{fig:2} top-middle). We then allow the vertex positions to relax to an energy minima (Fig.~\ref{fig:2}top-middle), and record the change in tissue energy $\delta E$ before and after the deformation. The shear modulus is then numerically estimated by $G = \frac{1}{A(0)}\frac{2 \delta E}{\epsilon^2}$.

\begin{figure}
    \centering
    \includegraphics[width=0.5\textwidth]{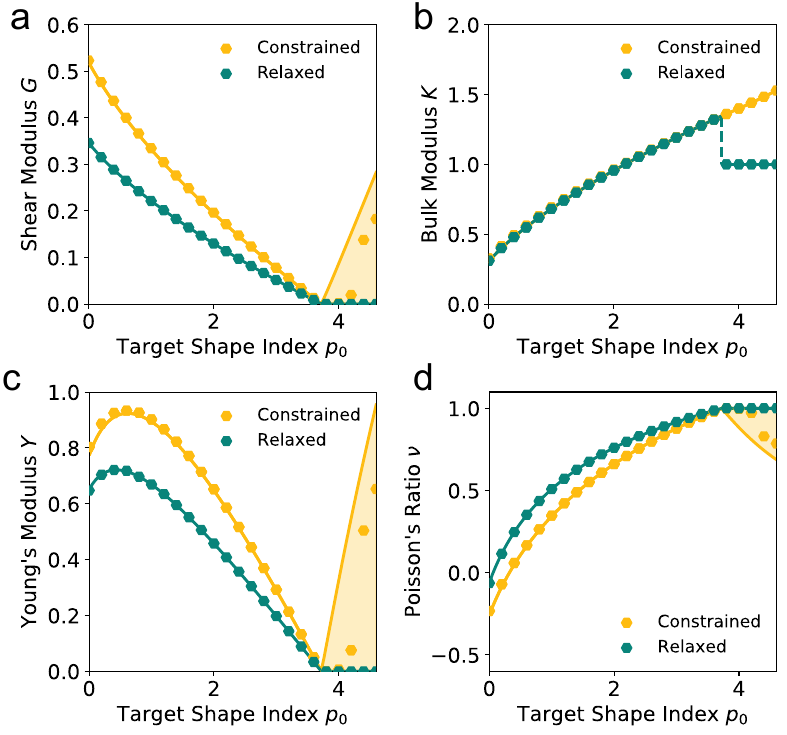}
    \caption{
    Non-affine deformations allow for a softer mechanical response. (a) Shear modulus $G$, (b) bulk modulus $K$, (c) Young's modulus $Y$, and (d) Poisson's ratio $\nu$ against target shape index $p_0$ for a rigidity ratio $r = 0.1$. The constrained values represent elastic moduli where vertices are constrained by the given deformation. The relaxed values lines represent the moduli allowing for non-affine deformations, where vertices may relax, subject to the boundary conditions. Dots represent simulated values. Lines represent analytic values. Shaded regions show the range of possible values in the constrained case, depending on the initial shape of the cells.}
    \label{fig:4}
\end{figure}

In the ground state of the incompatible regime, cell edges are under tension and meet at 120$^\circ$ angles. After the initial affine deformation, the angles change and the tissue is no longer in a force-balanced configuration (Fig.~\ref{fig:2} top-middle). As we allow the tissue to relax, it responds with a non-affine deformation; vertices which are of the same y-coordinate alternate between moving left and moving right during relaxation, returning the angles between edges to a stable 120$^\circ$ configuration (Fig.~\ref{fig:2} top-right). Such a deformation cannot be described by a single affine transformation, but by two affine transformations applied to different subsets of vertices~\cite{murisic2015discrete}.

To demonstrate the importance of this relaxation step, we report the response to two types of deformation protocols: (i) ``constrained'' deformations which are obtained where after deformation of the bounding box the cell vertices are not allowed to move to minimise the energy of the tissue, and (ii) ``relaxed'' deformations where the vertices are allowed to adjust their position to achieve force balance and  the global tissue shape remains controlled by the geometry of the deformed box. Note that in the compatible regime the relaxed state can also be achieved by allowing the tissue to change its shape~\cite{hernandez2022anomalous}, and the resulting linear elastic constants are the same.

For an intermediate rigidity ratio $r = 0.1$, we find that the shear modulus decreases as $p_0$ increases and becomes zero at the transition to the compatible regime. In particular, the relaxation step allows cells to decrease their perimeter, and thus energy, resulting in a relaxed shear modulus that is softer than in the constrained case (Fig.\ref{fig:4}a). In the compatible regime, the tissue is initially under no tension since the preferred perimeter is achieved. Upon straining the tissue, the perimeter increases and tissue energy increases. The larger the initial perimeter, the higher the change, resulting in constrained shear modulus that increases with $p_0$. When the tissue is able to relax, the vertices move to reduce the perimeter until the preferred perimeter is achieved again, allowing for the net energy to remain constant, leading to a zero shear modulus.

To calculate the bulk modulus, we apply the isotropic transformation $x_i(\epsilon) = (1 + \epsilon)^\frac{1}{2} x_i(0) $, $y_i(\epsilon) = (1 + \epsilon)^\frac{1}{2} y_i(0) $, and $L_x(\epsilon) = (1 + \epsilon)^\frac{1}{2} L_x(0)$ and $L_y(\epsilon) = (1 + \epsilon)^\frac{1}{2} L_y(0)$, where $\epsilon = 0.001$, such that  $A(\epsilon) = (1 + \epsilon)A(0)$. During the relaxation step, we allow the vertices to move, with the box lengths fixed. The bulk modulus is then given by $K = \frac{1}{A(0)}\frac{2 \delta E}{\epsilon^2}$.

In the incompatible regime, force balance requires a constant 120$^\circ$ angle between edges, thus the tissue expands isotropically. We find that the bulk modulus increases as the target shape index $p_0$ increases, and is equal between the relaxed and constrained cases (Fig.\ref{fig:4}b).

In the compatible regime, the deformation initially increases the perimeter. During the relaxation step, the tissue responds in a non-affine way to restore its perimeter to its preferred value and so energy change only arises from the area term and we have a bulk modulus $K = 1$. Interestingly, this is lower than the bulk modulus in the incompatible regime just before the transition and thus, there is a discontinuity in the bulk modulus as $p_0$ changes. In contrast, the constrained case is unable to relax the cells perimeters and so has a higher bulk modulus and does not exhibit the discontinuity (Fig.\ref{fig:4}b).

Next, we apply a uniaxial deformation to calculate the Young's modulus and Poisson's ratio: $x_i(\epsilon) = (1 + \epsilon)x_i(0)$ and $L_x(\epsilon) = (1 + \epsilon) L_x(0)$ (Fig.~\ref{fig:1}c). We then allow the vertex positions and box height $L_y(\epsilon)$ to relax to minimise energy. The Young's modulus is given by $Y = \frac{1}{A(0)} \frac{2 \delta E}{\epsilon^2}$ and the Poisson's ratio by $\nu = -\frac{(L_y(\epsilon) - L_y(0)) / L_y(0)}{(L_x(\epsilon) - L_x(0)) / L_x(0)}$. Note that this definition of the Poison’s ratio is equivalent to that in 2D elasticity and therefore its values are limited between $-1 < \nu < 1$. The extreme case $\nu=1$ corresponds to incompressible solid, analogous to the case of $\nu_{3d} = 0.5$ for incompressible 3D solids.

Again, the tissue undergoes a similar non-affine relaxation as under shear strain, reducing the shear modulus compared to the constrained case (Fig.~\ref{fig:4}c). In this case, though, we find that the Young's modulus is non-monotonic. For $p_0$ close to zero, the Young's modulus increases as $p_0$ increases. For higher $p_0$, increasing $p_0$ further decreases the Young's modulus towards zero at the transition point, after which the Young's modulus is zero. However, in the constrained case the Young's modulus increases after the transition point due to the increased bulk and shear moduli. Interestingly, the Poisson's ratio begins negative for small $p_0$ and increases towards a value of 1 as $p_0$ increases, before remaining 1 in the compatible regime (Fig.~\ref{fig:4}c). In the constrained case, the Poisson's ratio is actually lower than in the relaxed case for small $p_0$, in particular, in the compatible case the Poisson's ratio decreases as $p_0$ increases while for a relaxed tissue it remains $1$.

This phenomena highlights the counter-intuitive nature of VM mechanics. In classical elasticity $\nu = 1$ corresponds to incompressible solids, commonly considered as very stiff. Here we find that the tissue approaches $\nu = 1$ for higher values of $p_0$ corresponding to compatible tissue with floppy response. This seeming contradiction is resolved by noting that in this limit cells can accommodate rest area and perimeter simultaneously and therefore  upon deformation their area remains intact, just as in incompressible solids.

The simulations highlight the complex mechanical behaviour of the vertex model to applied tissue-level strains, both in its elastic moduli and the vertex-level non-affine deformations while relaxing the energy. The non-affine relaxation step allows the tissue to reduce its elastic moduli in the compatible regime. In particular, while the shear and Young's moduli are zero at the transition point, increasing $p_0$ increases the moduli in the constrained case, while they remain zero in the relaxed case. However, the simulations do not give an intuitive understanding for why the bulk modulus is discontinuous, or why we can get a negative Poisson's ratio. Thus, in the remainder of the paper, we develop a mean-field theory of the vertex model that can account for non-affine relaxation of the tissue under strain to derive analytic expressions for the elastic moduli and understand the source of the complex phenomena mentioned above.


\section{Vertex model: mean-field theory and ground states}
\label{sec:MFT}

\subsection{Mean-field theory of vertex model}
To understand the numerical results, we construct a mean-field theory by assuming that all cells responds equally. In this case the tissue energy is just $E_{\text{tissue}}=NE$ and one can simply consider the energy $E$ of a single cell, given by
\begin{equation}
   E = \frac{1}{2}(a - 1)^2 + \frac{1}{2}r(p - p_0)^2.
\end{equation}

\begin{figure}
    \centering
    \includegraphics[width=0.5\textwidth]{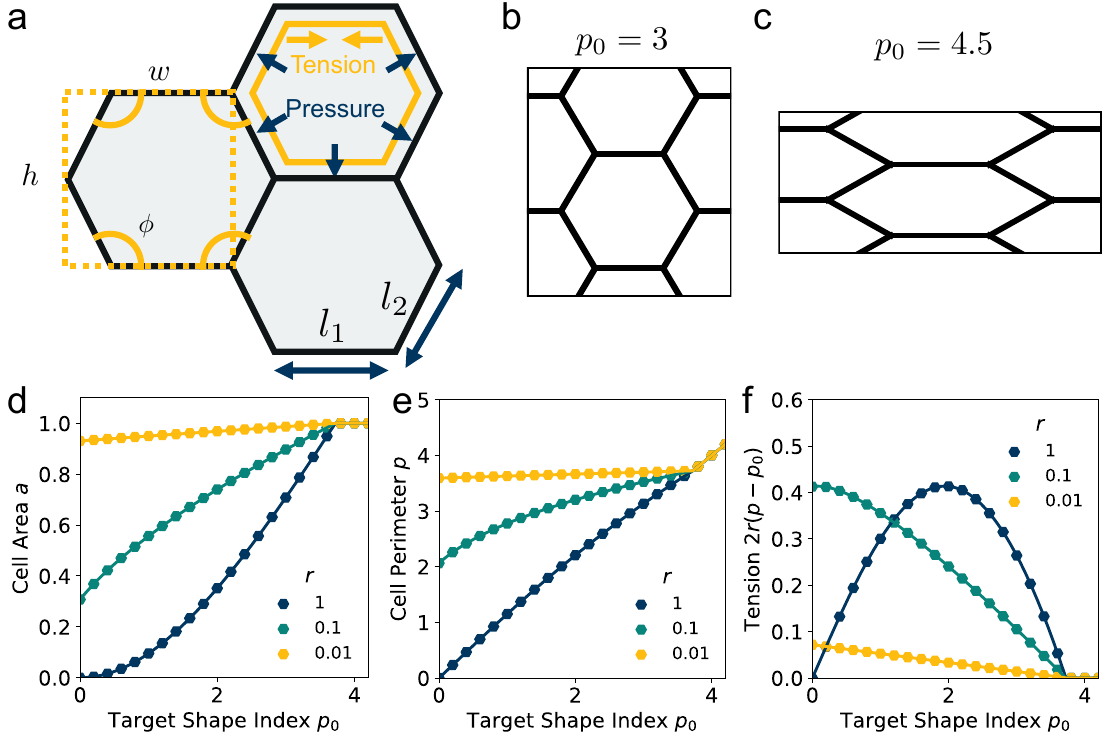}
    \caption{Shape parameterization of the vertex model and ground states. (a) Schematic of the vertex model and cell shape parametrization. Cells are defined by the lattice height $h$, width $w$, and angle between edges $\phi$. (b) The ground state in the solid state is a regular hexagonal lattice, with $\phi = 2 \pi / 3$. (c) The ground state used in the soft state with $\phi > 2 \pi / 3$. (d - f) Cell area (d),  cell perimeter (e), and  edge tension (f) vs target shape index $p_0$ for various values of the rigidity ratio $r$. Dots represent simulated values, lines are the  analytical results.}
    \label{fig:3}
\end{figure}

Each cell consists of horizontal edges of length $l_1$ and diagonal edges of length $l_2$, with  $\phi$ the angle between horizontal and diagonal edges (Fig.~\ref{fig:3}a). This parameterization captures the behavior of the tissue observed in our numerical simulations, where it is the angle between edges that changes during relaxation. Although cells have additional degrees of freedom, the description in terms of these three degrees of freedom is sufficient to capture the ground states of the tissue VM, and the response of the tissue under shear and bulk deformations in simulations (Fig.~\ref{fig:2}). To both examine the ground states and the response to deformation, it is convenient to parametrize each cell in terms of the height $h$ and width $w$, as shown in Fig.~\ref{fig:3}a, given by
\begin{equation}
    h = 2 l_2 \sin \phi\;,\hspace{0.2in}
 w = l_1 - l_2 \cos \phi\;.
   \label{eq:h-w}
\end{equation}
Each cell then contributes  an area $a = w h$ to the tissue.  We stress that the angle $\phi$ is distinct from the shear tilt angle $\theta$ introduced previously in Ref.~\cite{hernandez2022anomalous}, where cells may tilt or untilt in order to change their perimeter. While the shapes obtained from an initial regular hexagon by varying the shear tilt angle $\theta$ correspond to affine deformations of the hexagon, those parametrized by $\phi$ generally correspond to non-affine deformations of the regular hexagon.
Inverting Eqs.~(\ref{eq:h-w}), we obtain
\begin{align}
& l_1 = w + \frac{1}{2} h \cot \phi\;,\\
& l_2 = \frac{1}{2} h \csc \phi\;.
\end{align}
Cell area and perimeter can then be written as
\begin{align}
   & a = hw\;,\\
   & p = 2w + h f(\phi)\;,
\end{align}
where 
\begin{equation}
  f(\phi)=\frac{ 2
   + \cos \phi}{ \sin\phi}\;,
  \label{eq:fphi}
\end{equation} 
resulting in an energy
\begin{equation}
    E = \frac{1}{2}(hw - 1)^2 + \frac{1}{2}r(2w +  h f( \phi) - p_0)^2\;.
    \label{eq:Ephi}
\end{equation}
This form makes it evident that the VM energy is underconstrained as area and perimeter do not uniquely determine cell shape.

\subsection{Ground states}

The ground state configurations are obtained by minimizing the energy with respect to the cell width $w$, height $h$, and angle $\phi$ and are solutions of the three coupled equations
\begin{eqnarray}
\label{eq:dEdphi}
    \frac{\partial E}{\partial \phi} &=& h r f'(\phi)(2 w + h f(\phi) - p_0) = 0\;,\\
\label{eq:dEdh}
    \frac{\partial E}{\partial h} &=& w(hw - 1) + rf(\phi)(2 w + h f(\phi) - p_0) = 0\;,\\
\label{eq:dEdw}
    \frac{\partial E}{\partial w} &=& h(hw - 1) + 2r(2 w + h f(\phi) - p_0) = 0\;.
\end{eqnarray}
As shown in previous work, we find a transition at $p_0=p_*$ between two distinct states. For a regular lattice of $n$-sided polygons $p_*$ is given by the isoperimetric value $ p_*(n) = \sqrt{4 n \tan(\pi / n)}$, with $p_*(6) = \sqrt{8 \sqrt{3}} \approx 3.72$. The isoperimetric inequality $p\geq p_*(n)$ provides a lower bound on the perimeter of a regular $n$-sided polygon for given area~\cite{osserman1978isoperimetric}. For $p_0 > p_*(n)$ the cell is in a geometrically compatible regime, where both preferred area and perimeter may be achieved, and the tissue has zero shear modulus~\cite{staple2010mechanics, murisic2015discrete, moshe2018geometric, huang2021shear} (Fig.~\ref{fig:1}b). For $p_0 < p_*(n)$ the cell is in an incompatible regime, where both preferred area and perimeter cannot be simultaneously satisfied, and the tissue behaves like a solid by resisting shear deformation~\cite{farhadifar2007influence, moshe2018geometric, merkel2019minimal, hernandez2022anomalous}. The corresponding ground state of the tissue is a lattice of identical hexagonal cells (Fig.~\ref{fig:1}b). As $p_0$ is further lowered the cell may become unstable and collapse to zero area and perimeter (Fig.~\ref{fig:1}b). Additionally, in a small range of parameters near the collapsing region more exotic ground states exist, with mixed lattices of square and octagonal, or dodecahedral and triangular cells providing lower energy than  hexagonal cells~\cite{staple2010mechanics}.

\subsubsection{Compatible State, $p_0 > p_*(6)$}

For $p_0 > p_*(6)$ Eqs.~(\ref{eq:dEdphi}) are identically solved by $hw = 1$ and $p=2w + hf(\phi) = p_0$, and the zero ground state energy vanishes (Fig.~\ref{fig:3}c-e). We refer to this situation as the compatible state. The ground state configuration is a family of 6-sided polygons parametrized by the angle $\phi$, with
\begin{equation}
    h = \frac{p_0 \pm \sqrt{p_0^2 - 8 f(\phi)}}{2f(\phi)}\;,
    \label{eq:h_comp}
\end{equation}
\begin{equation}
    w = \frac{p_0 \mp \sqrt{p_0^2 - 8 f(\phi)}}{4}\;,
    \label{eq:w_comp}
\end{equation}
where both roots are acceptable solutions for a given value of $\phi$, corresponding to either tall and thin or short and wide cells. It is evident from Eq.~\eqref{eq:h_comp} and Eq.~\eqref{eq:w_comp} that 
such a solution exists provided $p_0^2 \ge 8 f(\phi)$. The function $f(\phi)$ has a minimum  at $\phi = \frac{2 \pi}{3}$, with $f(\frac{2 \pi}{3})=\sqrt{3}$ corresponding to $p_0 = 2^\frac{3}{2} 3^\frac{1}{4} = p_*(6)$. At this value of $p_0$ there is a single zero energy solution that corresponds to a hexagon of unit area. For $p_0>p_*$ there is degenerate continuum of zero energy solutions corresponding to deformed hexagons of unit area, perimeter $p_0$ and  $\phi\in\left[\frac{2 \pi}{3},\phi_m(p_0)\right]$, with $\phi_m$ determined by 
$p_0^2 = 8 f(\phi_m)$ (Fig.~\ref{fig:3}c). There exist many other parameterizations that can give ground state shapes in the compatible regime, for example, cells becoming tall and thin, cells decreasing the angle $\phi$ to increase their perimeter, or cells tilting as in Ref.~\cite{hernandez2022anomalous}.

\subsubsection{Incompatible State, $p_0 < p_*(6)$}

For $p_0 < p_*$  the cell cannot simultaneously realize the target area and perimeter. We refer to this situation as the incompatible state. Eq.~\ref{eq:dEdphi} requires $f'(\phi) = 0$, with solution $\phi = \frac{2 \pi}{3}$, such that $f(\phi) = \sqrt{3}$. An intuitive explanation for this fixed angle is that all edges are under identical tension, and so by force balance a junction of three edges must have equally spaced angles. Eqs.~\eqref{eq:dEdh} and~\eqref{eq:dEdw} then imply that $\sqrt{3} h = 2 w$. This gives a perimeter $p = 4w$ and area $a = \frac{2}{\sqrt{3}} w^2= p^2 / 8 \sqrt{3}$, which means that the cells are regular hexagons in the incompatible state (Fig.~\ref{fig:3}b).

We can combine Eqs.~\eqref{eq:dEdh} and~\eqref{eq:dEdw} to obtain a cubic equation for the perimeter
\begin{equation}
    p^3 + \left(\frac{r p_*^4 - 2 p_*^2}{2} \right) p - \frac{r p_*^4 }{2}p_0 = 0\;,
    \label{eq:p_min}
\end{equation}
with $p_*^2=8\sqrt{3}$ and $a=p^2/p_*^2$. The cubic equation can be solved perturbatively in the limit of low and high rigidity ratio $r$.

At low rigidity ratio, i.e., $r \ll 1 / p_*^2$, we find
\begin{align}
    p &= p_* - p_*^2 \left(\frac{p_* - p_0}{4}\right) r + \mathcal{O}(r^2)\;,\\
    a &= 1 -p_* \left(\frac{p_* - p_0}{2}\right)r + \mathcal{O}(r^2)\;.
\end{align}
The cell remains close in shape to a hexagon of unit area, with a reduction of the perimeter relative to the value $p_*$ (Fig.~\ref{fig:3}d-e). The tension, given by $2 r (p - p_0) = 2 r \left(p_* - p_0 \right) + \mathcal{O}(r^2)$, decreases monotonically as $p_0$ increases and vanishes at $p_0 = p_*$, where the cell reaches the compatible state and is under no tension  (Fig.~\ref{fig:3}f). If $p_0$ becomes too small, the stable configuration collapses to a point with zero area and perimeter (Fig.~\ref{fig:1}b).

For high rigidity ratio, i.e., $r \gg 1 / p_*^2$, the perimeter and area can be expanded in inverse powers of $r$, with the result
\begin{align}
    p &= p_0 + \frac{2 p_0}{p_*^2} \left(1 - \frac{p_0^2}{p_*^2} \right)\frac{1}{r} + \mathcal{O}\left(\frac{1}{r^2}\right)\;,\\
    a &= \frac{p_0^2}{p_*^2} + \frac{4 p_0^2}{p_*^4} \left(1 - \frac{p_0^2}{p_*^2}\right) \frac{1}{r} + \mathcal{O}\left(\frac{1}{r^2}\right)\;.
\end{align}
In this limit the cell shape index is close to the target shape index (Fig.~\ref{fig:3}d-e). The tension $2 r (p - p_0) = \frac{4 p_0}{p_*^2} \left(1 - \frac{p_0^2}{p_*^2} \right) + \mathcal{O}\left(\frac{1}{r}\right)$ is nonmonotonic in $p_0$, increasing from almost no tension at $p_0 = 0$ before decreasing as $p_0$ approaches $p_*$ (Fig.~\ref{fig:3}f).
As the rigidity ratio increases the tension saturates to the value  $\frac{4 p_0}{p_*^2} \left(1 - \frac{p_0^2}{p_*^2} \right)$, which has a maximum value of $2^\frac{3}{2} 3^{-\frac{7}{4}} \approx 0.414$ at $p_0 = 8 / \sqrt{3} \approx 2.149$ (Fig.~\ref{fig:3}d). For $p_0 \leq 0$, the cell collapses to a point with zero area and perimeter (Fig.~\ref{fig:3}b).

Finally, we note that when topological transitions are allowed, tissues may also unjam and undergo a solid-to-liquid phase transition when cell rearrangements cost zero energy. In disordered realizations of the VM, the unjamming transition occurs at $p_0 \approx 3.81$, a value close to, but slightly larger than $p_*(5)$~\cite{bi2015density}. Intuitively, for the tissue to rearrange in a T1 transition with zero energy barrier, two hexagonal cells must momentarily lose an edge and become pentagons while still maintaining their preferred perimeter and area. Cell motility can further promote fluidity and lower the transition point ~\cite{bi2016motility}.

\section{Mechanical response of the vertex model}
\label{sec:mechanics}

\subsection{Deformations protocol}

It is evident from Eq.~(\ref{eq:Ephi}) that area and perimeter (or equivalently height and width of the box shown in Fig.~\ref{fig:3}a) do not uniquely specify a polygonal shape. In the compatible regime there is a family of zero energy shapes corresponding to either tilted polygonal shapes obtained by affine deformations or non-affinely deformed polygons  parametrized by the angle $\phi$. In the incompatible regime, if only affine deformations are allowed, both the ground state and each deformed state are unique for fixed area and perimeter. Allowing non-affine deformations introduces, however, additional degrees of freedom that can lower the energy for a given set of parameters.

In the following we examine the response of a tissue initially in a ground state to an externally imposed  strain. The deformation is imposed globally on the tissue by changing the shape of the bounding box. Such a deformation uniformly changes the shape of the cells and generally results in a state where individual vertices are no longer force balanced (Fig.~\ref{fig:2} top-middle). We will refer to this state as the ``constrained'' deformed state. Due to the presence of hidden degrees of freedom the system can, however, lower its energy and relax a state of local force balance. In the compatible regime this relaxation can occur via motion of the vertices that correspond to non-affine deformations (Fig.~\ref{fig:2} top-right). In the compatible regime the relaxation can occur either via non-affine deformations with fixed box shape or through a global tilting of the tissue, which entails affine cell deformations, as in Hernandez et al.~\cite{hernandez2022anomalous}. The elastic constants measured in the ``relaxed'' state of the compatible regime are the same for the two relaxation protocols.

Operationally, constrained deformations are achieved by first fixing either the cell height, width, or both, and then transforming the vertices according to the given deformation, as done in Staple \emph{et al.}~\cite{staple2010mechanics}. We prevent spontaneous tilting of the tissue, which can be used to soften the mechanical response using only affine deformations in the compatible regime~\cite{hernandez2022anomalous}.

We next evaluate the various elastic moduli of the vertex model. As we will see below, a new result of our work is that in the incompatible regime cells can find new deformed states by relaxing through non-affine deformations (Fig.~\ref{fig:2}), resulting in a softer response than obtained in previous studies~\cite{staple2010mechanics} (Fig.~\ref{fig:4}).

\subsection{Shear Modulus}

To calculate the shear modulus of the tissue, we apply an area-preserving pure shear deformation, corresponding to $w \rightarrow w (1 + \epsilon / 2)$ and $h \rightarrow h (1 + \epsilon / 2)^{-1}$, with $\epsilon$ the strain. We allow for a non-affine deformation to relax the tissue by minimising energy with respect to the angle $\phi$ for each value of strain (Fig.~\ref{fig:3}). The shear modulus is defined as
\begin{equation}
    G = \frac{1}{a} \left. \frac{\partial^2 }{\partial \epsilon^2}\left( \min_\phi E\right) \right |_{\epsilon=0}\;.
    \label{eq:shear}
\end{equation}
As area is preserved under pure shear, we  only need to consider the energy cost due to changes in perimeter.

\subsubsection{Compatible State, $p_0 > p_*$}

\begin{figure}
    \centering
    \includegraphics[width=0.5\textwidth]{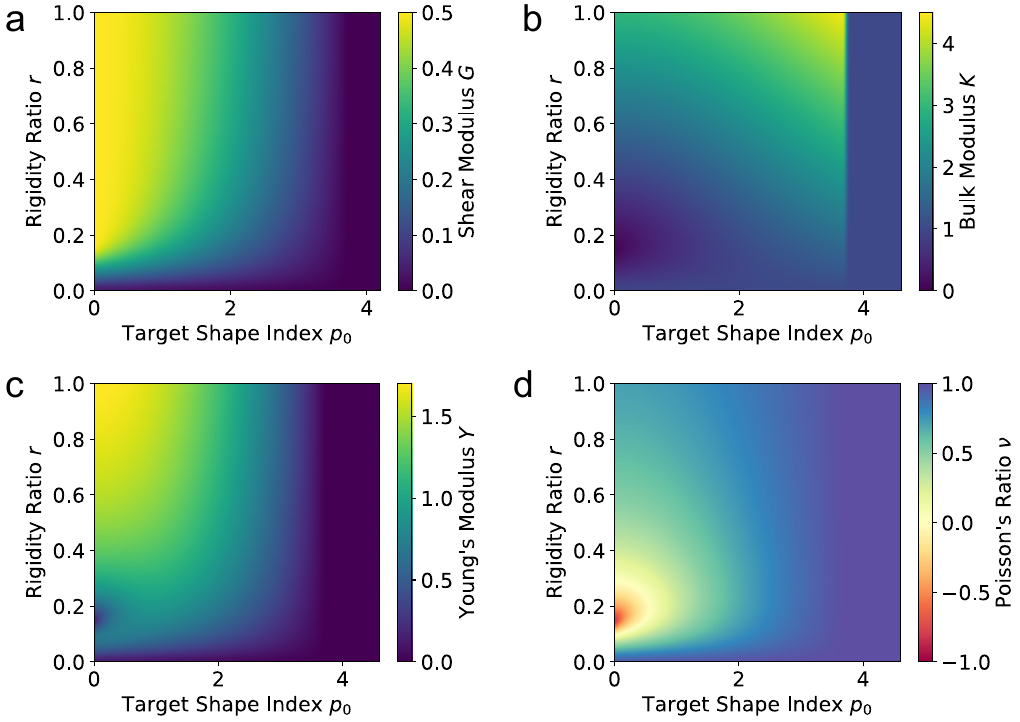}
    \caption{Elastic moduli of the vertex model. (a) Shear modulus, (b) bulk modulus, (c) Young's modulus, and (d) Poisson's ratio against target shape index $p_0$ and rigidity ratio $r$.}
    \label{fig:5}
\end{figure}

In the compatible case, cells can accommodate shear and maintain their area and perimeter at the target values $a=1$ and $p=p_0$ by changing shape, i.e., by adjusting the angle $\phi$ to a value other than $2\pi/3$. The perimeter of the deformed cell is given by $p(\epsilon,\phi)=2w(1+\epsilon / 2)+hf(\phi)/(1+\epsilon / 2)$. The cell can maintain $p=p_0$ by deforming to a new compatible ground state corresponding to an angle $\phi^*$ given by the solution of  $p(\epsilon,\phi^*)=p_0$. Clearly the energy remains zero, demonstrating that the shear deformation cost no energy and 
\begin{equation}
    G = 0
\end{equation}
for all rigidity ratios and $p_0 > p_*$ (Fig.~\ref{fig:5}a). This of course only holds up to a maximum value of strain determined by the angle $\phi_m(p_0)$. Beyond this value the fluid-like compatible tissue stiffens and acquires a finite shear modulus, as mentioned by \cite{farhadifar2009dynamics}, and recently by~\cite{huang2021shear}.

However, if the shape of the cell is constrained then applying a shear transformation could increase the perimeter. To calculate the shear modulus under an affine transformation, we must consider the changes to the cell perimeter from its initial configuration. The shear modulus is defined by
\begin{equation}
    G_{\text{affine}} = \frac{1}{a} \left. \frac{\partial^2 E}{\partial \epsilon^2}\right |_{\epsilon=0} = \frac{1}{a} r \left(\frac{\partial^2 p}{\partial \epsilon^2} (p - p_0) + \left(\frac{\partial p}{\partial \epsilon}\right)^2 \right)|_{\epsilon=0}\;.
    \label{eq:shear}
\end{equation}
and since for $\epsilon = 0$ we have $a = 1$ and $p = p_0$ this simplifies to
\begin{equation}
    G_{\text{affine}} = r  \left(\frac{\partial p}{\partial \epsilon}\right)^2|_{\epsilon=0}\;.
    \label{eq:shear}
\end{equation}
Since an affine transformation also changes the angle between the edges, we must consider the effect of the transformation on each edge when calculating the change in perimeter. The length of the two edges shown in Fig.~\ref{fig:3}a change as 
\begin{equation}
l_1(\epsilon) = l_1(0) \left(1 + \frac{\epsilon}{2} \right)
\end{equation}
and
\begin{equation}
    l_2(\epsilon) = l_2(0) \left(\left(1 + \frac{\epsilon}{2} \right)^2 \cos^2 \phi + \left(1 + \frac{\epsilon}{2} \right)^{-2} \sin^2 \phi \right)^\frac{1}{2};
\end{equation}
with first derivatives
\begin{equation}
    \frac{\partial l_1}{\partial \epsilon} = \frac{1}{2} l_1(0)
\end{equation}
and
\begin{equation}
    \frac{\partial l_2}{\partial \epsilon} = \frac{1}{2} l_2(0) \frac{\left(1 + \frac{\epsilon}{2} \right) \cos^2 \phi - \left(1 + \frac{\epsilon}{2} \right)^{-3} \sin^2 \phi }{\left(\left(1 + \frac{\epsilon}{2} \right)^2 \cos^2 \phi + \left(1 + \frac{\epsilon}{2} \right)^{-2} \sin^2 \phi \right)^\frac{1}{2}};.
\end{equation}
Thus
\begin{equation}
    \frac{\partial p}{\partial \epsilon} |_{\epsilon = 0} = l_1(0) + 2 l_2(0) (\cos ^2 \phi - \sin^2 \phi)
\end{equation}
and
\begin{equation}
    G_{\text{affine}} = r \left(l_1(0) + 2 l_2(0) (\cos ^2 \phi - \sin^2 \phi) \right)^2\;.
\end{equation}

The cell angle $\phi$ defines a family of solutions with $\sqrt{3} < f(\phi) < p_0^* / 8$, as does the choice between the $\pm$ branch in our solution for cell height and, and so we obtain a range of values of the constrained shear modulus, and thus Young's modulus and Poisson's ratio, as $\phi$ is varied. We find that as $p_0$ is increased, the minimum shear modulus in the constrained case remains $0$ while the maximum possible shear modulus increases (Fig.~\ref{fig:4}a).

\subsubsection{Incompatible Case, $p_0 < p_*$}

In the incompatible case cell edges are under uniform tension. By force balance, this implies that the angle between edges remains $\phi = \frac{2 \pi}{3}$ even under small deformations at the tissue scale. The ground state configuration is a regular hexagon  with perimeter $p = 2w + \sqrt{3}h$. Using Eq.~\eqref{eq:shear} and the relations for height and width in terms of the perimeter, $h = \frac{1}{2\sqrt{3}}p$ and $w = \frac{1}{4}p$, we obtain
\begin{equation}
    G =\frac{r(p - p_0)p}{4a}\;.
\end{equation}
In the rigid, incompatible state the shear modulus is a monotonically increasing function of $r$ and vanishes at the transition $p_0 = p_*$ (Fig.~\ref{fig:4}a, Fig.~\ref{fig:5}a), in agreement with earlier results~\cite{murisic2015discrete}. The non-affine deformations of the relaxed tissue allow for a softer response of the tissue, with the shear modulus being a factor of $3/2$ stiffer when only considering vertices constrained by the affine shear strain~\cite{farhadifar2009dynamics, staple2010mechanics}, as confirmed by simulations (Fig.~\ref{fig:4}a).

For small rigidity ratio ($r \ll 1 / p_*^2$), we can expand $G$ in powers of $r$, with the result
\begin{equation}
G =\frac{1}{4}p_{*}(p_{*}-p_0) r + \mathcal{O}(r^2)\;.
\end{equation}
The opposite limit of large rigidity ratio ($r \gg 1 / p_*^2$)  yields
\begin{equation}
G=\frac{1}{2}\left(1-\frac{p_0^2}{p_*^2}\right)-\frac{1}{p_*^6}\left(p_*^2-p_0^2\right)^2\frac{1}{r} + \mathcal{O}\left(\frac{1}{r^2}\right)\;.
\end{equation}

\subsection{Bulk Modulus}

To calculate the bulk modulus, we change the area $a \rightarrow a(1 + \epsilon)$ by rescaling the height $h \rightarrow h(1 + \epsilon)^\frac{1}{2}$ and width $w \rightarrow w(1 + \epsilon)^\frac{1}{2}$, and allow the angle $\phi$ to vary to minimize the deformation energy. The bulk modulus is then given by
\begin{equation}
    K = \frac{1}{a}\left.\frac{\partial^2 }{\partial \epsilon^2}\left( \min_{\phi} E\right)\right|_{\epsilon=0}\;,
\end{equation}
with
\begin{equation}
\frac{\partial^2 E}{\partial \epsilon^2} = a^2 + r(p - p_0)\frac{\partial^2 p}{\partial \epsilon^2} + r\left(\frac{\partial p}{\partial \epsilon}\right)^2\;.
\end{equation}
To evaluate this expression we need to consider separately the compatible and incompatible states.

\subsubsection{Compatible state, $p_0>p_*$}
We have previously shown that in the compatible case the angle $\phi$ can adjust to maintain a fixed cell perimeter under small deformations. Thus $\left. \frac{\partial p}{\partial \epsilon} \right|_{\epsilon=0} = 0$ and $\left. \frac{\partial^2 p}{\partial \epsilon^2} \right|_{\epsilon=0} = 0$, and the bulk modulus is simply 
\begin{equation}
    K = 1
\end{equation}
for all $r$ and $p_0 > p_*$ (Fig.~\ref{fig:4}b, Fig.~\ref{fig:5}b).

By contrast, if we allow for only affine deformations then the perimeter expands isotropically $p(\epsilon) = (1 + \epsilon)^\frac{1}{2} p(0)$. In this case $\left. \frac{\partial p}{\partial \epsilon} \right|_{\epsilon=0} = \frac{1}{2}p(0)$ and $\left. \frac{\partial^2 p}{\partial \epsilon^2} \right|_{\epsilon=0} = -\frac{1}{4}p(0)$, giving a bulk modulus equal to
\begin{equation}
    K_{\text{affine}} = 1 + \frac{1}{4} r p_0^2,
\end{equation}
which can be significantly higher than the non-affine result for high $r$, emphasising the need to consider non-affine displacements (Fig.~\ref{fig:2}).

\subsubsection{Incompatible State, $p_0<p_*$}

In the incompatible state for $p_0 < p_*$  the angle that minimizes energy remains $\phi = \frac{2 \pi}{3}$ for small perturbations to cell height and width to ensure tension balance  at the cell vertices. The cell then expands isotropically, such that $p(\epsilon) = (1 + \epsilon)^\frac{1}{2} p(0)$, resulting in a bulk modulus

\begin{equation}
    K = a + \frac{1}{4a}r p p_0\:
\end{equation}
shown in Fig.~\ref{fig:4}b and Fig.~\ref{fig:5}b. We note that as $p_0$ approaches the critical value $p_*(6)$ from below,  the bulk modulus has the value $\rm{lim}_{p_0\rightarrow p_*^-}K = 1 + \frac{1}{4}r p_*^2$. On the other hand, in the compatible regime $K = 1$. Thus the bulk modulus exhibits a jump discontinuity at the critical point separating compatible and incompatible states. In contrast, if vertex positions are fixed by uniform dilation without relaxation, the bulk modulus is continuous and higher in the compatible region (Fig.~\ref{fig:4}b).

In the limit of low rigidity ratio  ($r \ll 1 / p_*^2$), we find
\begin{equation}
    K = 1 + \frac{1}{4}p_* (3 p_0 - 2 p_*) r + \mathcal{O}(r^2)\;.
\end{equation}
The bulk modulus increases with $p_0$ up to the critical value, at which point it discontinuously jumps to $1$ for all $p_0 > p_*$, independent of $r$. Interestingly, for $p_0 < \frac{2}{3} p_*$ the bulk modulus of the incompatible solid is lower than that of the compatible fluid, suggesting that contractility can actually reduce the bulk stiffness of the tissue. Additionally, increasing the rigidity ratio further reduces the bulk modulus for low $p_0$.

In the limit of high rigidity ratio ($r \gg 1 / p_*^2$), we find
\begin{equation}
    K = \frac{1}{4} p_*^2 r + \left(\frac{3}{2} \frac{p_0^2}{p_*^2} - \frac{1}{2} \right) + \mathcal{O}\left(\frac{1}{r}\right)\;.
\end{equation}
Thus the bulk modulus increases with the rigidity ratio.

We find that for low $p_0$ the bulk modulus can be significantly lower in the rigid than in the fluid state, which can affect the rate of spreading monolayers~\cite{staddon2022interplay}. This is due to the fact that at small  $p_0$ cells have a smaller area while the energy required to deform the cell is proportional to the square of the area change. Therefore it costs more energy to strain a single cell than to strain two cells with half the area, similar to the reduction in effective stiffness obtained when springs are placed in series.

The bulk modulus is also a non-monotonic function of  the rigidity ratio. For small $r$ the bulk modulus decreases with $r$ as the size of the cell decreases, reaching a minimum near $r = 2 / p_*^2 \approx 0.144$ before increasing linearly in $r$ for high $r$, due to the growing contribution from the perimeter elasticity.

\subsection{Young's Modulus and Poisson's Ratio}

Next we calculate the Young's modulus and Poisson's ratio (Fig.~\ref{fig:2} bottom row) by stretching the width of the cell $w \rightarrow w ( 1 + \epsilon)$ while allowing the cell height $h$ and angle $\phi$ free to minimize the energy. The Young's modulus is defined as 
\begin{equation}
    Y = \frac{1}{a} \left. \frac{\partial^2 }{\partial \epsilon^2}\left(\min_{h,\phi}  E\right) \right|_{\epsilon = 0}
\end{equation}
and the Poisson's ratio as
\begin{equation}
    \nu = -\frac{1}{h} \frac{\partial h}{\partial \epsilon}= -\frac{w}{h} \frac{\partial h}{\partial w}.
\end{equation}
To evaluate $Y$ and $\nu$ we use the relationship between the linear elastic constants, $Y = \frac{4KG}{K + G}$ and $\nu = \frac{K - G}{K + G}$, which have been shown to hold away from the critical point.

\subsubsection{Compatible state, $p_0 > p_*$}

In the compatible state, the ground state degeneracy allows the cell to achieve the target shape index and area for small strain by reducing cell height and finding values of the angle $\phi$ different from $2\pi/3$, i.e., by changing its shape, with no energetic cost. As a result for $p_0 > p_*$ we find
\begin{equation}
    Y = 0\;,~~~~~
    \nu = 1\;.
\end{equation}
for all $r$ (Fig.~\ref{fig:4}c-d, Fig.~\ref{fig:5}c-d). When constrained to affine only deformations, the Young's modulus can take a range of values from zero to a maximum value which increases with $p_0$, due to the increasing shear modulus. Similarly, the Poisson's can range from a maximum of 1 to a minimum value which decreases with $p_0$.

\subsubsection{Incompatible case, $p_0 < p_*$}

It has been shown that away from the critical point the elastic constants of the VM satisfy the familiar relation of linear elasticity of isotropic solids~\cite{hernandez2022anomalous}. We can therefore use the relations $Y = \frac{4KG}{K + G}$ and $\nu = \frac{K - G}{K + G}$ to evaluate $Y$ and $\nu$ in the incompatible regime, with the result
\begin{align}
    Y &=(p-p_0) \frac{rp\left(4a^2 + rp p_0 \right)}{a(4a^2 + rp^2)}
    \;,\\
    \nu &= 1-\frac{2rp(p-p_0)}{4a^2+rp^2}
    \;.
\end{align}
We find that the Young's modulus is a non-monotonic function of both target shape index and rigidity ratio (Fig.~\ref{fig:4}c, Fig.~\ref{fig:5}c). The nonmonotonicity with $p_0$ is most pronounced at intermediate values of $r$, where at small $p_0$ the Young's modulus increases, rather than decrease, with increasing $p_0$.

At both high and low rigidity ratio, the Poisson's ratio remains close to $1$ for all $p_0$, indicating that the cell preserves its area under deformations (Fig.~\ref{fig:4}d, Fig.~\ref{fig:5}d). At intermediate values of the rigidity ratio and small $p_0$, the nonmonotonicity of the Young's modulus results in a negative Poisson's ratio, which indicates that  a tissue stretched in the $x$-direction, also expands in the $y$-direction.

In comparison to the constrained response, we find that the relaxed response gives a softer Young's modulus for all rigidity ratio and target shape index values, since the shear modulus is also softer by a factor of $\frac{2}{3}$ (Fig.~\ref{fig:4}c). Similarly, the Poisson's ratio is higher in the compatible state for all $p_0 < p_*$ (Fig.~\ref{fig:4}d).

In the limit of low rigidity ratio ($r \ll 1 / p_*^2$), the Young's modulus and Poisson's ratio are given by
\begin{align}
    Y &= p_* (p_* - p_0) r + \mathcal{O}(r^2)\;,\\
    \nu &= 1 + \frac{1}{2} p_* (p_0 - p_*) r + \mathcal{O}(r^2)\;,
\end{align}
showing that when  $p_0$ is increased towards the critical point from the solid side ($p_0\rightarrow p_*^-$) the Young's modulus vanishes and the Poisson's ratio increases towards $1$.

In the limit of high rigidity ratio ($r \ll 1 / p_*^2$), we obtain approximate expression by expanding in $1/r$ as
\begin{align}
    Y &= 2 \left(\frac{p_*^2 - p_0^2}{p_*^2}\right) + \mathcal{O}\left(\frac{1}{r}\right)\;,\\
    \nu &= 1 -\frac{4 (p_*^2 - p_0^2)}{p_*^4}\frac{1}{r} + \mathcal{O}\left(\frac{1}{r^2}\right)\;.
\end{align}
The Young's modulus also has a maximum value of $2$ at $p_0=0$.

\subsection{Origin of negative Poisson's ratio}

We can understand why certain cell parameters give a positive or negative Poisson's ratio by looking at how the energy gradient with respect to cell height changes as we change the cell width. The gradient $\frac{\partial E}{\partial h}$ can be thought of as the effective force acting on the height of the cell, given by
\begin{equation}
    \frac{\partial E}{\partial h} = w(hw - 1) + \sqrt{3}r(2w + \sqrt{3} h - p_0).
\end{equation}
In the ground state, this will be zero. Then, if we vary the cell width but keep the height fixed we can measure the change in force as
\begin{equation}
    \frac{\partial^2 E}{\partial w \partial h} = 2 h w - 1 + 2 \sqrt{3} r.
\end{equation}
When this value is positive, then as cell width is increased, the effective force acting on cell height increases and so the cell height will decrease as it relaxes to the energy minimum. Consequently, the sign of this value is the same sign as the Poisson's ratio.

The second term $2\sqrt{3}r$ comes from the perimeter contribution in the VM and accounts for energy changes due to perimeter elasticity. Since $p \geq p_0$, the cell is under tension and the perimeter term aims to shrink the cell. Increasing the width further increases the perimeter and so tension increases, providing more force to shrink the cell. Thus, the perimeter elasticity always acts to shrink the cell and contributes to a positive Poisson's ratio.

The first term, $2hw - 1 = 2a - 1$, represents the energy change due to area elasticity. We can write this as $2w (h - 1 / 2w)$, which we can think of as an spring like force with stiffness $2w$ and target height $1/2w$. As width increases, the height becomes closer to the target height, reducing the strain. At the same time, the effective stiffness $2w$ increases, increasing the pressure. Thus there is a trade off between less restoring force on the cell area versus increased effectiveness of changes in cell height. The net effect on whether this increases or decreases the perimeter depends on the size of the cell area: when cell area $a < \frac{1}{2}$ the area term acts to increase cell height when width is increased, and for $a > \frac{1}{2}$ the area term reduces the cell height.

For high rigidity ratio, the perimeter term dominates and so an increase in cell width leads to a reduction in cell height. For low rigidity ratio, the area term dominates and the cell area is close to $1$ (Fig.~\ref{fig:3}d), thus an increase in cell width reduces the area pressure and cell height decreases. However, in the intermediate regime for low $p_0$ cell area is small, meaning the area term acts to increase cell width, and the area contribution and perimeter contributions are of comparable size, resulting in negative Poisson's ratio.

We can calculate the transition to a negative Poisson's ratio exactly at  $p_0 = 0$, which corresponds to the situation where the contribution to cell edge tension from cortical contractility and cell-cell adhesion precisely balance. In this limit the equation for the ground state perimeter, Eq.~\ref{eq:p_min}, becomes
\begin{equation}
    p \left(p^2 + \frac{r p_*^4 - 2 p_*^2}{2} \right) = 0
\end{equation}
with  solution 
\begin{align}
p &= p_* \sqrt{1 - \frac{1}{2} r p_*^2}\;.\\
    a &= \left(1 - \frac{1}{2}r p_*^2 \right)
\end{align}
for $r < 2 / p_*^2 \approx 0.144$. For $r > 2 / p_*^2$ the cell is unstable and collapses to zero area. We can calculate whether cell height increases or decreases when width is increased by calculating how the effective force on the height, $\frac{\partial E}{\partial h}$, changes with width. Substituting our formula for area we find
\begin{equation}
    \frac{\partial^2 E}{\partial h \partial w} = 1 - \frac{3}{4} r p_*^2= 1 - 6\sqrt{3} r
\end{equation}
where we have used $p_*^2 = 8 \sqrt{3}$. Thus for $r > \frac{4}{3}p_*^2 \approx 0.096$ and $p_0 = 0$ the tissue has a negative Poisson's ratio. 

\section{Conclusions}
\label{sec:conclusions}

In this paper we studied the linear response of the 2D vertex model by calculating the shear modulus, bulk modulus, Young's modulus and Poisson's ratio, using a mean-field approach that allows for a class of non-affine deformations, which agree well with numerical simulations. We also provide approximate expressions in the limit of high and low rigidity ratio. Our calculations match previous results showing a rigidity transition controlled by purely geometric effects and tuned by the target shape index $p_0$.

For cells in the incompatible case, $p_0 < p_* \approx 3.772$, the tissue has a finite shear modulus which decreases with $p_0$. For cells in the compatible case, $p_0 > p_*$, the shear modulus becomes zero for all $p_0$. However, when the tissue is constrained by the deformation, we find a stiffer mechanical response to shear in the incompatible case, and the compatible case can have a finite shear modulus, which depends on the initial configuration of the cells, and that increases with $p_0$.

The bulk modulus of the tissue increases with $p_0$ in the incompatible regime, and then has a jump discontinuity at $p_0 = p_*$, where it changes from a larger value in the solid state to a value of $1$ in the fluid state. In the incompatible case, cells perimeters increase upon isotropic expansion of the tissue, but in the compatible regime cells can change shape to preserve their perimeter under small changes in area. However, this discontinuity is not observed when the tissue is constrained. We also find that in the incompatible regime, the bulk modulus can decrease below $1$ for small rigidity ratio. This indicates that cell contractility can reduce the stiffness of the tissue, resulting in a larger bulk modulus in the ``soft'' phase than in the ``solid'' phase.

Under uniaxial strain, we find the Young's modulus of the tissue can be non-monotonic with respect to $p_0$, initially increasing and then decreasing towards zero in the incompatible case. The Poisson's ratio can become negative for small $p_0$ and intermediate rigidity ratio, as cells can reduce their energy more by increasing their area through an orthogonal expansion, than by reducing their perimeter. Within the compatible regime, the tissue has zero Young's modulus and Poisson's ratio equal to one. However, when only constrained deformations are allowed the tissue can have a finite Young's modulus in the compatible regime, similar to the shear modulus.

Our results highlight the complex linear elastic behaviour that can arise from the simplest version of the vertex model due to its underconstrained nature. For simplicity, we have assumed that cells are regularly arranged and that we only have small strains. This analysis might be most applicable in tissues with a regular crystalline structure, such as the \textit{Drosophila} pupal wing~\cite{classen2005hexagonal.} However, it would be interesting to extend our calculations to the case of disordered cell networks. Additionally, we highlight the importance of allowing for unconstrained degrees of freedom, in this case non-affine deformations, to relax the system and give a softer mechanical response to strain. The constrained case may be thought of as the short-time response of the tissue to strain, and the relaxed case as the long-time limit.

Finally, we note that the emergence of rigidity observed in this work draws a direct link with the rigidity of mechanical frames and granular models, where mechanical stability occurs at critical coordination number, or at finite strains, and are normally accompanied by a discontinuous jump in the bulk modulus. Nevertheless, the results of our work shows that geometric incompatibility is a crucial ingredient that has to be taken into account for an estimation of the critical coordination number, and is left for a future work~\cite{damavandi2022energetic1, damavandi2022energetic2}.

In conclusion, our work demonstrates that the vertex model, thought of as a collection of geometric constraints rather than a reference ground state structure, can engender interesting linear mechanical responses. The linear response exhibits a strong non-affine contribution under uniaxial compression and shear, as well as a negative Poisson's ratio. Typically, these two phenomena in crystalline solids require special lattice constructions, whereas in the vertex model exotic mechanical response can be achieved by tuning the relative competition between area and perimeter constraints via $r$ and geometric compatibility via $p_0$.


\section*{Conflicts of interest}
There are no conflicts to declare.

\section*{Acknowledgements}
M.C.M. and A.H. were supported by the National Science Foundation Grant No.~DMR-2041459. M.M. was supported by the Israel Science Foundation grant No. 1441/19. This research was supported in part by the National Science Foundation under Grant No. NSF PHY-1748958.




\bibliography{arxiv_update} 

\end{document}